\documentclass[12pt]{article}
\usepackage[dvips]{graphicx}
\usepackage[T2A]{fontenc}
\usepackage[utf8]{inputenc}
\usepackage{natbib,url}
\begin{document}

\title{Carbon Supported Polyaniline \\ as Anode Catalyst: \\
       Pathway to Platinum-Free Fuel Cells}

\author{A. G. Zabrodskii, M. E. Kompan, 
V. G. Malyshkin\footnote{mal@poly-aniline.com} \\
Ioffe Physico-Technical Institute, RAS \\
St. Petersburg, 194021 Russia \\
I. Yu. Sapurina\footnote{sapurina@hq.macro.ru} \\
Institute of Macromolecular Compounds, RAS \\
St. Petersburg, 199004 Russia
}%

\date{April 11, 2006}

\maketitle
\begin{abstract}
The effectiveness of carbon supported polyaniline as anode catalyst in a fuel cell (FC) with direct
formic acid electrooxidation is experimentally demonstrated. A prototype FC with such a platinum-free composite
anode exhibited a maximum room-temperature specific power of about 5 $mW/cm^2$

PACS: 82.47.Gh   82.65.+r   82.45.Fk


\end{abstract}

 1. The main advantage of fuel cells
(FC) employing liquid organic fuels as compared to
those using hydrogen is the simplicity of storage and
recharging. Among the FC with liquid fuels, the most
thoroughly studied are FC with direct methanol electrooxidation 
(DMFC) [1]. The other potential liquid organic
fuels include formic acid. In FC employing direct 
formic acid electrooxidation (DFAFC) [2], a somewhat
lower energy capacity (as compared to that of DMFC)
is compensated by other advantages, including 
relatively simple reaction nature (and, accordingly, 
facilitated problem of catalyst selection) and lower extent of
fuel crossover through the proton-conducting (nafion)
membrane.

    As is known, direct electrochemical oxidation of
organic fuels on the FC anode leads to the formation of
 $CO_2$ (complete oxidation). In the case of DFAFC, the
process is:

\begin{eqnarray}
HCOOH&\rightarrow&CO_2 +2H^{+} +2e^{-}
\label{rand}
\end{eqnarray}

\noindent
The active evolution of carbon dioxide is a reliable
manifestation of this reaction. The process on the 
cathode consists in the reduction of molecular oxygen $O_2$  with
the formation of water$H_2O$ :

\begin{eqnarray}
\frac{1}{2} O_2 + 2H^{+} +2e^{-} &\rightarrow& H_2O
\label{rcat}
\end{eqnarray}

\noindent
It should be noted that cathodic reaction (2) is common
to all FC with proton-conducting membranes, which
use oxygen as the oxidizer.

    The main problems inherent in all FC with direct
electrochemical oxidation of organic fuels are the
following:

    (a) The output power yield is limited by the rate of
anodic reaction and usually varies at room temperature
within 5---20 $mW/cm^2$ [3--5].

    (b) The fuel crossover through a nafion membrane to
cathode and its oxidation on the FC cathode surface is
equivalent to an opposite emf [4, 6], that is, to a
decrease in output power and in the FC efficiency.

    (c) Some organic fuels produce degradation of
nafion membrane in the course of FC operation.

   The anode and cathode of FCs are electrodes with
catalytic coatings, which are usually based on platinum
(Pt). The limited resource of this metal on the Earth
stimulates the interest in creating nonplatinum catalysts.

   For cathodic reaction (2), catalysts containing no
metallic platinum have been developed based on 
macrocyclic complexes (see [3, 7] and references therein),
biological materials [8], and transition metal 
compounds [3, 6, 7]. It was also reported that polyaniline
(PANI) exhibited catalytic activity in the reaction of
oxygen reduction in air--metal cells [9]. All these 
systems exhibit a lower catalytic activity than platinum,
but their development and practical use nevertheless
good have prospects.

   In recent years, FC units have been developed with
extensive use of conducting polymers (synthetic metals)
such as PANI, polypyrrole, and polythiophene [10--12].
Some data [3, 13] and our previous results [14, 15] 
indicate that polymers with electron--proton conductivity
(such as PANI) can increase the efficacy of 
platinum group metal catalysts [16].

   As for anodic reaction (1), no catalysts free of 
platinum group metals have yet been reported, although
there were communications on the catalytic activity of
PANI in the reactions of anodic oxidation of hydrogen
[17], methanol [18], and ascorbic acid [19]. As is
known, PANI combines a high level of electron (hole)
conductivity ($1$ --- $10$ S/cm [11, 12]) with proton 
conductivity (up to $10^{-2}$ S/cm) [15, 20, 21]. The mixed 
conductivity type of this polymer is of key importance in
electrochemical processes involving simultaneous
transport of both protons and electrons. It should also
be noted that PANI has a variable structure; this 
polymer contains benzoid and quinoid fragments linked via
nitrogen atoms occurring in various oxidation and 
protonation states. The ratio of the fragments of various
types can change in a reversible manner depending at
least on two parameters of the reaction medium: 
oxidation potential and acidity [11]. This circumstance 
provides broad possibilities for controlled modification of
the properties of PANI.

2. This preprint
reports the process of electrochemical oxidation of 
formic acid in an FC with the anode made of a carbon
material coated with a nanolayer (100 nm) of PANI.
PANI in an emeraldine form was obtained via aniline
oxidation with ammonium peroxydisulfate 
immediately on the surface of fibers of a porous carbon 
material [22]. The resulting PANI was strongly adsorbed on
the carbon substrate. The polymer nanolayer 
encapsulated carbon fibers in the entire volume of the porous
carbon matrix, so that the total weight fraction of PANI
in the resulting composite reached 20\%.

   The electro-
chemical cell schematically depicted in Fig. 1. In order
to study the anodic reaction, we used a cathode 
half--MEA (membrane--electrode assembly) [23] 
comprising a Nafion 117 membrane with the standard DMFC
cathode containing $\rm 4mg/cm^2$ of platinum. The cathode
operated in air under natural convection conditions.

    The anode was made of Torray TGPH-060 carbon
cardboard covered with PANI nanolayer as described
above. The cardboard was pressed against the nafion
membrane with a thicker cardboard 
(free of PANI coating), which simultaneously served 
as electrode and gasdiffusion medium. The fuel was a 5\% solution of 
formic acid (HCOOH) in a 0.5M aqueous $H_2SO_4$ solution.
Since the experiments were performed with a fuel 
possessing ionic conductivity, the problem 
of contact quality in the anodic region was not so important, 
since protons could readily pass from one medium to another via
the ion-containing fuel.

    Figure 2 shows the experimental loading 
characteristics measured using a prototype FC describe above.
Plots of the FC output voltage versus current density in
the membrane had the typical form, beginning at about
0.7V (at a nearly zero current density) and rapidly
decaying at current densities below 5 $mA/cm^2$ . Using
the subsequent less steep decrease, the internal 
resistance of the FC prototype was estimated at about
10  $\Omega cm^2$. The maximum specific output power
reached in our experiments at room temperature was
about 5 $mW/cm^2$ . This value corresponds to the output
power of the typical DMFCs operating at room 
temperature (20C) [3, 4]. In the region of high current 
densities (10--40 $mA/cm^2$), the curves were poorly 
reproduced in different experimental runs.

  As was noted above, the active gas evolution on the
FC anode operating on formic acid fuel is a reliable 
criterion for reaction (1) to actually take place. Indeed,
when the FC was connected to a low-ohmic load and a
relatively large current passed through the cell, 
intensive evolution of $CO_2$ bubbles (Fig. 3) was observed.
The volume of liberated gas was measured and 
compared to the amount of charge transported vie the FC
circuit. In this calculation, we assumed that the 
oxidation of the HCOOH molecule yields a charge of 
$2\times 1.6 \cdot 10^{-19} C$. 
The volume of $CO_2$ calculated for the
transferred charge was 1.8 times the volume evolved in
the experiment. The reason for so large a discrepancy is
still unclear. One possible explanation is offered by the
following mechanism: a fraction of current could be, in
principle, related to the additional oxidation of PANI
with the formation of pernigraniline. However, 
calculations showed that an additional charge provided by
complete oxidation of PANI present on the anode was
two orders of magnitude lower than the total charge
transported in the experiment.

    Thus, the electrochemical oxidation of formic acid
is the only process that can be responsible for the 
liberation of energy in the prototype FC studied. Moreover,
if the observed current were related to PANI oxidation,
the output current would unavoidably drop from one
run to another, which was not observed in our experiments.

    The experiments showed no systematic decrease in
the output power during the first two days. In the family
of loading curves presented in Fig. 2, the curves 
corresponding to the maximum current and density were
obtained in the last experimental run of the series. The
current and output power exhibited a reversible
decrease as a result of the fuel consumption in the FC
and were restored on the initial level upon adding a new
portion of the fuel. However, the experiments 
performed in the following days showed a decrease in the
current and approximately proportional decrease in the
FC open-circuit voltage (down to 0.4V), and even to a
lower level in the subsequent week. We explain this
behavior by the diffusion (crossover) of formic acid in
the membrane, which results in the appearance of fuel
on the oxidizer side and is equivalent to the opposite
emf operation [4, 6]; an additional detrimental factor is
degradation of the nafion membrane surface in contact
with formic acid.

    It is necessary to emphasize the stability of results.
The electrochemical oxidation of formic acid is not
characteristic of the given type of carbon material. 
Specific output power on a level of $3-5 mW/cm^2$ was also
obtained in prototype FCs where PANI was supported
(instead of Torray TGPH-060) on carbon materials of
the Kinol ACC-10-20 or Busofit T-1-55 types.

   Thus, the results of our experiments convincingly
demonstrated the catalytic activity of PANI in the
anodic reaction of formic acid oxidation.

3. We can only suggest some notions 
concerning the nature of the catalytic activity of PANI. One
possible mechanism is the reduction of PANI from the
emeraldine to leuco-emeraldine form, which is 
accompanied by the oxidation of formic acid and is followed
by leuco-emeraldine oxidation to emeraldine and 
electron transfer to the anode. We believe that the redox
transition of PANI from the emeraldine form to a lower
oxidation state (leuco-emeraldine) mediates in the 
electron transfer and accelerates the oxidation of formic
acid.

    Another special question concerns the possible role
of the carbon-PANI interface in the enhancement of the
catalytic activity. We suggest that, since PANI is a
p-type conductor [24] and carbon materials possess
metallic conductivity, this interface features a potential
barrier of the Schottky type. Since PANI is a medium
permeable to liquids (in this case, to formic acid),
HCOOH molecules occur in a strong electric field near
the proton-conducting membrane and the field induces
their polarization, which can in principle lead to a
decrease in the dissociation energy. The close 
phenomenon of the polarization and subsequent ionization of
shallow impurity states in the electric field is well
known in semiconductor physics [25].

4. Thus, we have experimentally 
demonstrated that carbon supported PANI exhibits high 
catalytic activity in the reaction of anodic oxidation (1) of
formic acid in fuel cells of the DFAFC type. This 
activity was observed in a working prototype FC with 
nonplatinum composite anode of the carbon supported
PANI type, which ensured a stable specific output
power of about 5 $mW/cm^2$ over a long period of time.
In combination with published data on the nonplatinum
catalysts for the cathodic reaction, these results open the
way to the creation of FCs entirely free of platinum 
catalysts, which is the aim of our further investigations [26].

This study was supported in part
by the Presidential Program of Support for Leading 
Scientific Schools in Russia (project no. NSc-5920.2006.2),
the FANI Program (project no. 02.434.11.7054), and
the Russian Foundation for Basic Research (project
nos. 06-02-16991a and 04-02-16672a).

\newpage

\begin{figure}
\includegraphics[width=8cm]{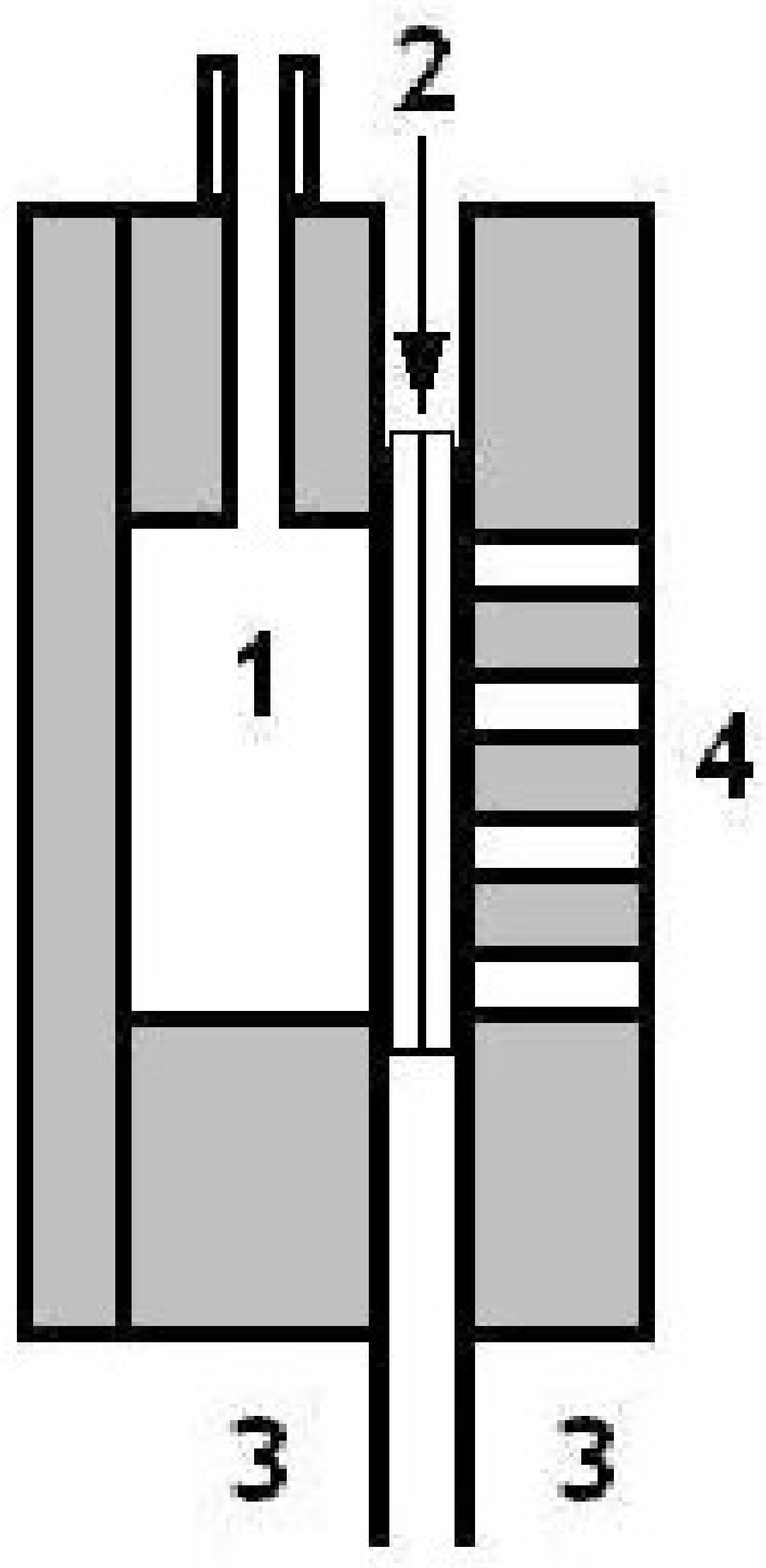}
\caption{\label{figcell}
Schematic diagram of the electrochemical cell
involving a cathode half-MEA: (1) liquid fuel container;
(2) membrane--electrode assembly (see the text for details);
(3) plate electrodes with holes; (4) case with air channels.
Spacers and screws are not shown.
}
\end{figure}

\begin{figure}
\includegraphics[width=16cm]{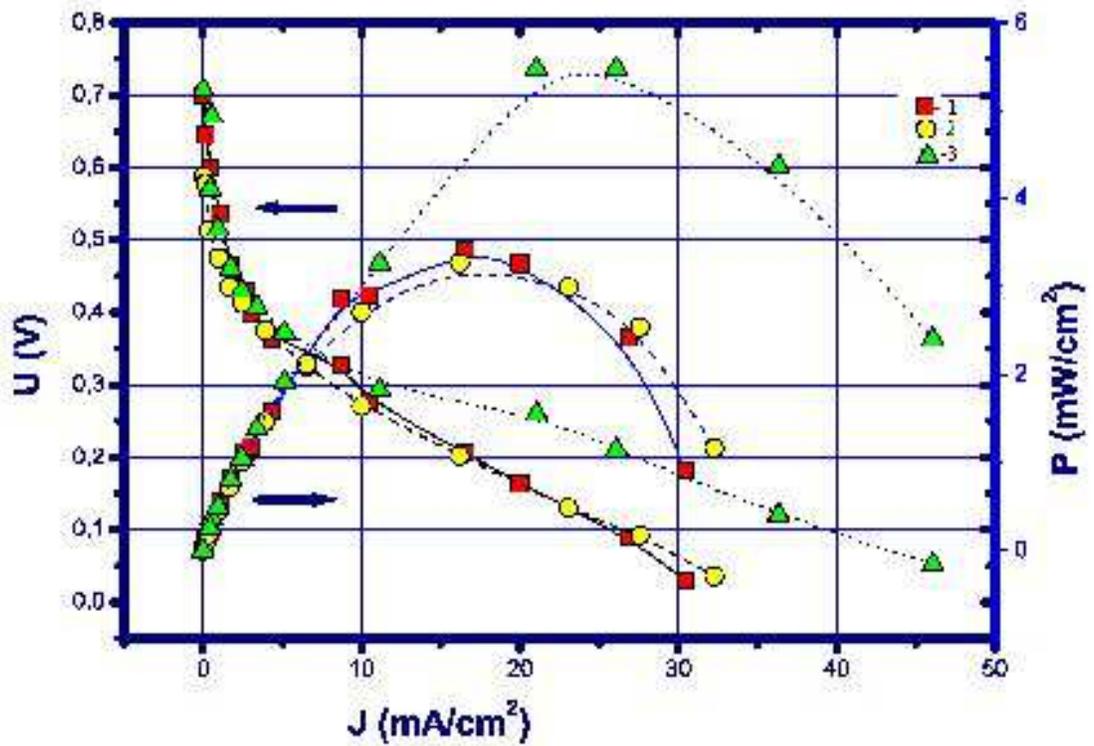}
\caption{\label{figvi}
 Plots of voltage U (left scale) and specific power
(right scale) versus current density measured in a prototype
FC. Curves 1--3 refer to three sequential experimental runs
performed with a 10-min time interval.
}
\end{figure}

\begin{figure}
\includegraphics[width=15cm]{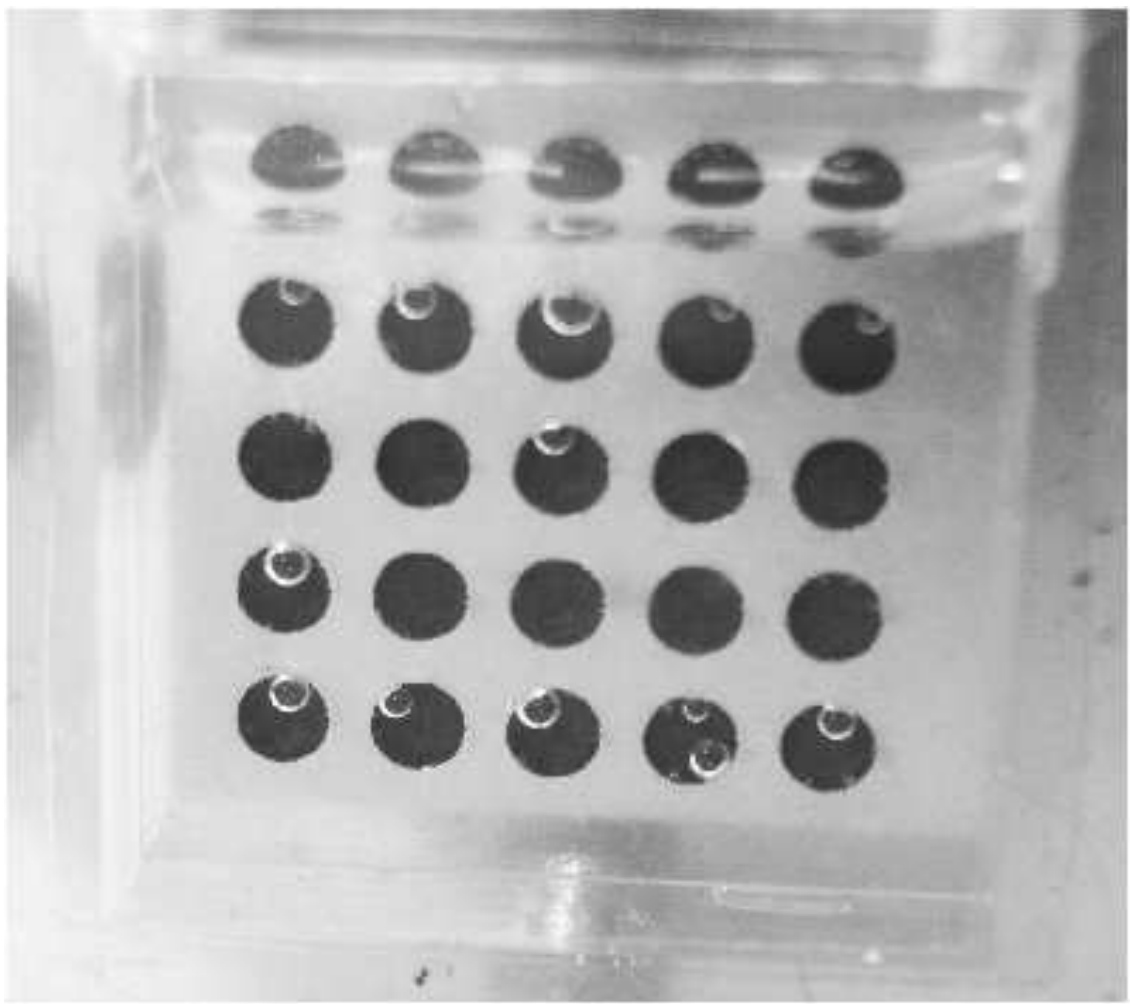}
\caption{\label{figpuzyri}  $CO_2$ evolution on the anode of operating FC.
}
\end{figure}

\end{document}